\documentclass[conference]{IEEEtran}
\IEEEoverridecommandlockouts
\usepackage{cite}
\usepackage{amsmath,amssymb,amsfonts}
\usepackage{algorithmic}
\usepackage{graphicx}
\usepackage{textcomp}
\usepackage{xcolor}

\def\BibTeX{{\rm B\kern-.05em{\sc i\kern-.025em b}\kern-.08em T\kern-.1667em\lower.7ex\hbox{E}\kern-.125emX}}
\begin{document}

\title{Predicting Bandwidth Utilization on Network Links Using Machine Learning
}

\author{\IEEEauthorblockN{Maxime Labonne}
\IEEEauthorblockA{\textit{Institut LIST, CEA} \\
F-91120, Palaiseau, France \\
maxime.labonne@cea.fr}
\and
\IEEEauthorblockN{Charalampos Chatzinakis}
\IEEEauthorblockA{\textit{Institut LIST, CEA} \\
F-91120, Palaiseau, France\\
charalampos.chatzinakis@cea.fr}
\and
\IEEEauthorblockN{Alexis Olivereau}
\IEEEauthorblockA{\textit{Institut LIST, CEA} \\
F-91120, Palaiseau, France\\
alexis.olivereau@cea.fr}
}

\maketitle

\begin{abstract}
Predicting the bandwidth utilization on network links can be extremely useful for detecting congestion in order to correct them before they occur. In this paper, we present a solution to predict the bandwidth utilization between different network links with a very high accuracy. A simulated network is created to collect data related to the performance of the network links on every interface. These data are processed and expanded with feature engineering in order to create a training set. We evaluate and compare three types of machine learning algorithms, namely ARIMA (AutoRegressive Integrated Moving Average), MLP (Multi Layer Perceptron) and LSTM (Long Short-Term Memory), in order to predict the future bandwidth consumption. The LSTM outperforms ARIMA and MLP with very accurate predictions, rarely exceeding a 3\% error (40\% for ARIMA and 20\% for the MLP). We then show that the proposed solution can be used in real time with a reaction managed by a Software-Defined Networking (SDN) platform.
\end{abstract}

\begin{IEEEkeywords}
Congestion detection, LSTM, MLP, ARIMA, Real-Time Bandwidth Prediction
\end{IEEEkeywords}

\section{Introduction}
All companies that offer network services (ISPs, server hosting services etc.) use mechanisms to monitor link utilization. These mechanisms usually involve network interface monitoring and collection of performance statistics (e.g. with SNMP \cite{Case:1990:SNM:RFC1157}), monitoring of flows (e.g. with NetFlow \cite{Sommer:2002:NIL:637201.637226}, sFlow \cite{Phaal:2001:ICS:RFC3176}, etc.) or capturing packets and further analyzing them with a specific tool \cite{SurveyMonitoring}. Detection of high network utilization is a problem that needs to be addressed efficiently since it usually causes packet loss, increased latency due to buffering of packets, and interference with TCP's congestion-avoidance algorithms. The result is a degradation of the Quality of Service (QoS) of the network.

There are numerous ways to circumvent the problem like allocating more bandwidth to accommodate for the increased traffic, prioritizing important traffic through QoS, blocking undesirable traffic, or load balancing the traffic across multiple paths \cite{doi:10.1002/047002531X.ch2}. However, all these solutions come after the bandwidth problem has been detected. Predicting the future bandwidth consumption would allow to proactively correct this problem.

One of the most important metrics for network performance evaluation is the network link utilization at any given time. Link utilization is usually expressed as a percentage of the total capacity of a network link. For example an 100 Mbps Ethernet link that carries 20 Mbps of traffic exhibits 20\% utilization. The higher the percentage of usage, the lower the quality of the link, resulting in packet loss and increased latency.

Our solution predicts network utilization using machine learning algorithms. A simulated network is created to collect data from network and system resource statistics. They are processed with feature engineering and merged to create a dataset, which is fed to the machine learning algorithms. Three models are tested: ARIMA, MLP and LSTM. Our goal is to identify weak signals among the features of the dataset to predict the bandwidth utilization.

The remainder of the paper is organized as follows. Section 2 discusses related work about congestion detection. Section 3 introduces the data generation process using a simulated network. Section 4 describes the preprocessing stage, from data collection to the creation of the dataset. Section 5 presents the machine learning models used to predict the bandwidth utilization on network links and our experimental results. Finally, section 6 discusses areas of future work and concludes this paper.

%%%%%%%%%%%%%%%%%%%%%%%%%%%%%%%%%%%%%%%%%%%%%%%%%%%%%%%%%%%%%%%%%%%%%%%%%%%%%%%%%%%%%%%%%%%%%%%%%%%%%%%%%%%%%%%%%

\section{Related Work}

Congestion detection is a useful tool to improve the performance of any type of network. Many solutions have been proposed at the network protocol level or at the application level. The use of machine learning algorithms has gained prominence over the last two decades in the literature.

Devare and Kumar \cite{Devare2007ClusteringAC} applied time series analysis to congestion control using clustering and classification techniques. They generated traffic patterns for a number of clients, selected relevant data and preprocessed them to build a database. Different machine learning classification algorithms are trained with these data and used to predict the future traffic intensity. Countermeasures are then applied to improve the performance of the network.

Singhal and Yadav \cite{congestion1} used neural networks to detect congestion in wireless sensor networks. The authors created their own dataset using NS-2 \cite{Issariyakul:2010:INS:1952077} to generate a random traffic. Three features are used as inputs for the machine learning algorithm: the number of participants, the traffic rate, and the buffer occupancy. This neural network has a 3-10-10-1 structure and predicts a level of congestion (low, medium, or high).

Madalgi and Kumar \cite{8389138} explored a similar idea with two congestion detection classifiers. Their dataset is also generated with NS-2 and machine learning algorithms are used to predict the same three levels of congestion. The authors compared the performance of a MLP and a decision tree (M5 model tree). They show that the decision tree is trained faster (0.53 second against 8.25 seconds) and outperforms the MLP in terms of accuracy, true positive rate, and false positive rate.

The same authors \cite{8768738} proposed another work using Support Vector Machines (SVMs). They specifically used LibSVM and Sequential Minimal Optimization (SMO), with Radial Basis Function (RBF) as kernel. Different parameters were tweaked to find the best values to improve the classification accuracy. The best model is trained in 18.81 seconds and achieves slightly lower results than the previous M5 decision tree.

Zhang et al. \cite{8854045} worked on a specific type of network congestion: low-rate denial of service attacks. This type of attacks exploits the TCP congestion-control mechanism to deplete the resources of the target. The authors used the Power Spectral Density (PSD) entropy function to reduce the number of calculations: a flow is classified as normal below a certain threshold, and as an attack above a second threshold. A SVM classifies uncertain connections between these two thresholds, using 8 features. The experimental results show that this solution can detect 99.19\% of the low-rate denial of service attacks in the dataset.

%%%%%%%%%%%%%%%%%%%%%%%%%%%%%%%%%%%%%%%%%%%%%%%%%%%%%%%%%%%%%%%%%%%%%%%%%%%%%%%%%%%%%%%%%%%%%%%%%%%%%%%%%%%%%%%%%

\section{Data Generation Using A Simulated Network}

\begin{figure}[htpb]
\centering \includegraphics[scale=0.25]{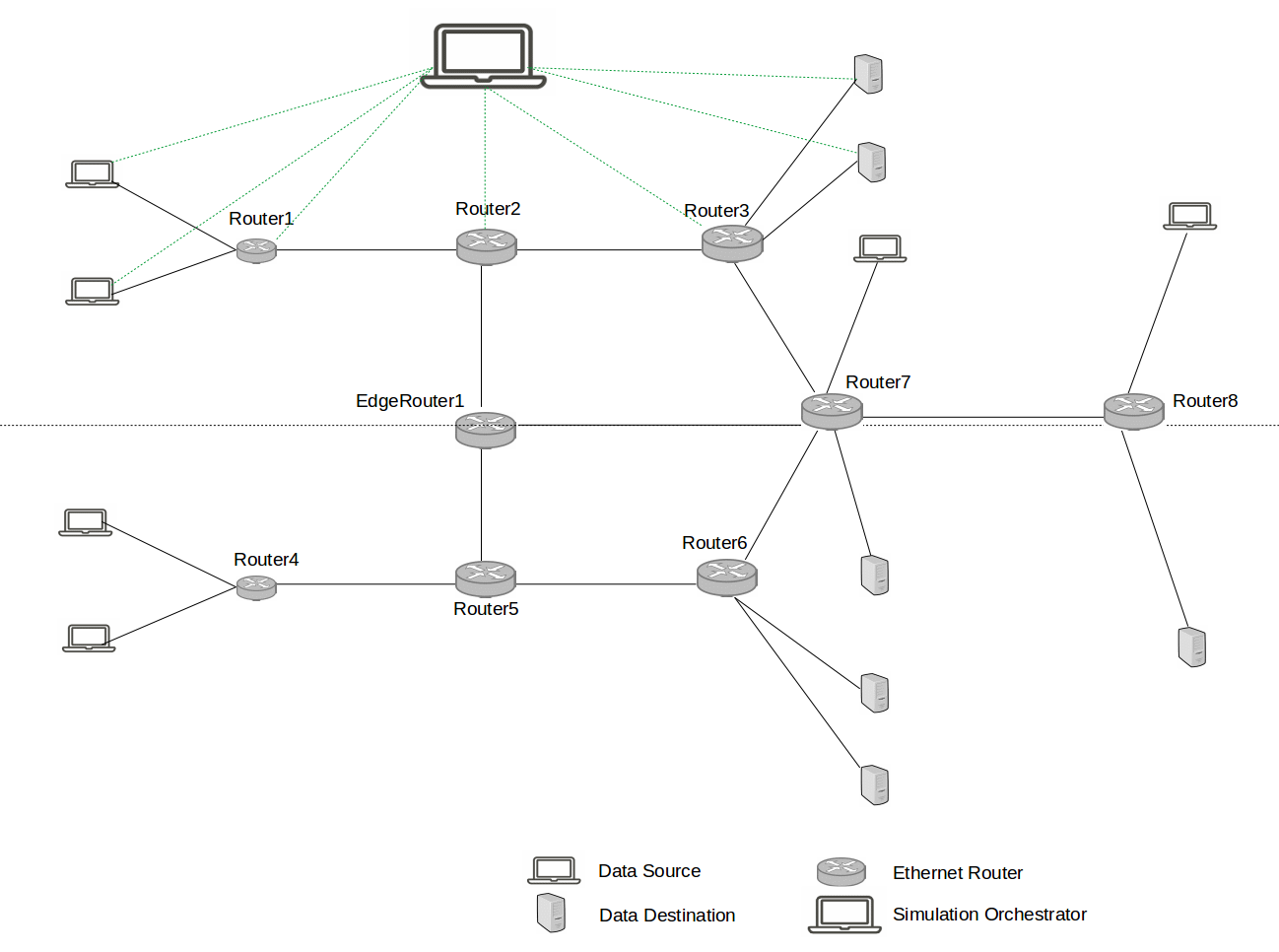}
\caption{Topography of the Simulated Network.}
\label{figure-topology}
\end{figure}

A testbed network was created in order to collect data for the machine learning algorithms. Regular TCP and UDP traffic was generated which occasionally caused heavily loaded links in order to simulate real-world scenarios. The network topology is illustrated in Fig. \ref{figure-topology}.

Virtualbox was used to simulate the network with each end host and router being a separate virtual machine. Network links were configured with a maximum capacity of 100 Mbps and with 2 ms of latency. This bandwidth capacity does not represent state-of-the-art data center switches, but is a good representation of our problem. Finally, a separate VM called Orchestrator was created to orchestrate the simulation phase on the network.

iPerf3 was used to simulate traffic in the network \cite{iperf}. iPerf3 is primarily used for bandwidth measurements of network links by generating TCP and UDP flows with certain parameters. Using iPerf3 as a traffic generator has its limitations as it cannot always generate realistic network traffic. For example, it mainly generates packets with fixed inter-departure times and cannot create bursty traffic. It is, however, easily scriptable and allowed us to quickly generate training data for our models. Moreover, we have randomized some of the flow parameters to introduce some diversity into the network traffic. The random flow parameters are summarized in Table \ref{tab:flow-table}.

\begin{table}[htpb]
\centering
\caption{Randomized Flow Parameters}
\label{tab:flow-table}
\begin{tabular}{|c|c|c|}
\hline
\textbf{Parameter} & \textbf{Values} & \textbf{Distribution} \\ \hline
Protocol & TCP, UDP & Uniform \\ \hline
Bitrate & \begin{tabular}[c]{@{}c@{}}Average: 2 Mbps\\ 			Std: 500Kbps\end{tabular} & Normal \\ \hline
Flow Duration & {[}30,500{]} seconds & Uniform \\ \hline
Packet Length & {[}200,1472{]} bytes & Uniform \\ \hline
Flow Inter-arrival Time & {[}20,2000{]} seconds & Uniform \\ \hline
Source/Destination couple & IP of the end hosts & Uniform \\ \hline
iPerf3 Server Port & {[}5001,5500{]} & Uniform \\ \hline
\end{tabular}
\end{table}

The bitrate, flow duration, and flow inter-arrival time intervals were chosen to ensure that maximum capacity would be reached on certain or all links, even for a short period of time. In other words, we had to make sure that we had data ranging from zero traffic to maximum capacity while keeping simulation times relatively short. Although real-world network traffic is primarily TCP and not UDP, we decided not to favor one over the other.

During the simulation phase, a script running on the Orchestrator was in charge of starting the data collection mechanism on each one of the routers. It started every $[20,2000]$ seconds an iPerf3 server on a destination machine and an iPerf3 client on a source machine, thus creating a new data flow with certain characteristics. The results presented later in this paper come from a three-day simulation, where all the interfaces of each one of the routers were monitored. The data were collected in the form of text files with a predetermined frequency (every 3 seconds). The simulation generated around 85\,000 samples per interface for a total of 2 million samples (12 GB) for all the 22 monitored interfaces.

%%%%%%%%%%%%%%%%%%%%%%%%%%%%%%%%%%%%%%%%%%%%%%%%%%%%%%%%%%%%%%%%%%%%%%%%%%%%%%%%%%%%%%%%%%%%%%%%%%%%%%%%%%%%%%%%%

\section{Preprocessing Stage}

\subsection{Data Collection}

Machine learning algorithms for supervised learning require labelled data as input. The prediction quality is directly related to the relevance of these data. Two tools were used in order to obtain a maximum of information on the bandwidth of the network links: Netmate and Dstat.

\textbf{Netmate} (Network Measuring and Accounting Meter) \cite{netmate} is a network measurement tool that can collect a number of network statistics, such as packet volumes and sizes, packet inter-arrival times, and flow duration. There are 46 features collected by Netmate for every active flow at the time of the capture: protocol (TCP, UDP), total forward and backward packets, total forward and backward volume, duration, distribution metrics (packet length, inter arrival time, active time, idle time). A separate instance of Netmate was run for every interface on a router. This tool was configured to export its output as a .csv file every 3 seconds for every interface. Every .csv file was timestamped with the seconds since epoch format.

\textbf{Dstat} \cite{dstat} is a versatile Linux tool for generating system resource statistics from numerous system components like CPU, RAM, input/output devices, networks connections and interfaces, and others. This tool is highly configurable and can output the collected data into a .csv file for further processing. Dstat collected 29 features with our configuration.

\subsection{Feature Engineering}

Data from Netmate and Dstat require a preprocessing phase in order to create datasets in a format suitable for machine learning algorithms. Given the fact that Netmate generates its statistics for every active flow, a feature engineering procedure is necessary to create new features that would describe the status of the link collectively. Mathematical functions like $min()$, $max()$, $mean()$, $sum()$ and $count()$ were applied to the Netmate data to obtain 86 new features. For instance, the total number of packets can be calculated with the sum of the number of packets for each of the active flows: $total\_pkt\_vol = sum(pkt\_volume\_of\_each\_flow$). These 86 features are then concatenated with the 29 features coming from Dstat using the timestamp value to synchronize the output of the two tools. After this feature engineering step, 116 features are collected from every interface of each router in our network every 3 seconds.

\begin{figure}[htpb]
\centering \includegraphics[scale=0.43]{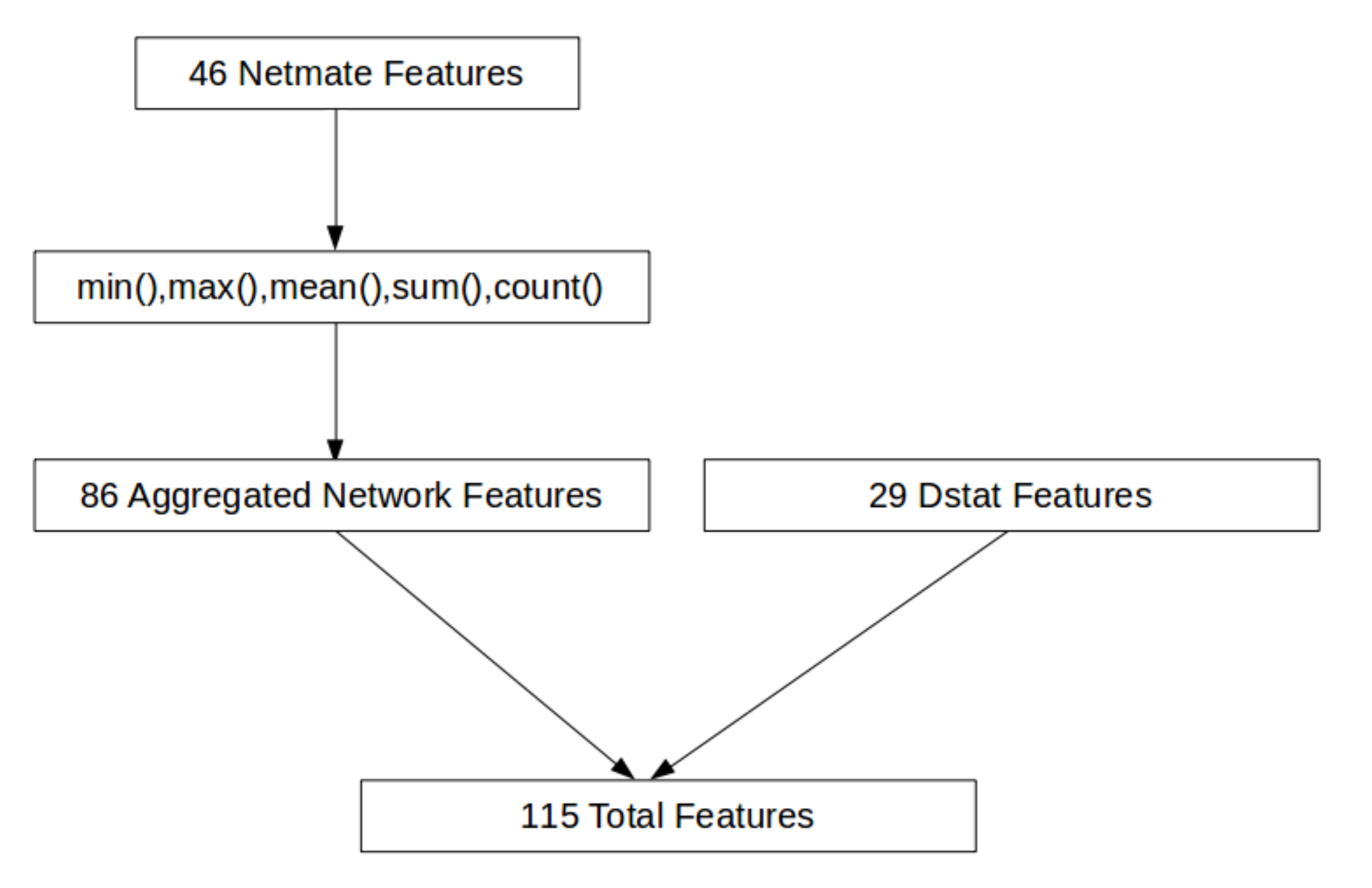}
\caption{Feature Engineering Workflow.}
\label{figure-feature-engineering}
\end{figure}

The feature engineering process is illustrated in Fig. \ref{figure-feature-engineering}.

\subsection{Creation of a Dataset}

A dataset for supervised learning needs to be labeled for the training process. The two most important features to predict the link usage are the $download\_bitrate$ and $upload\_bitrate$ provided by Dstat. A new feature called $max\_bitrate$ is created by choosing the maximum of the two values and dividing it by the nominal maximum capacity of the link (100 Mbps). In other words, the new column gives the link usage ratio, either in the uplink or the downlink direction. The $max\_bitrate$ column is then duplicated to create the $future\_bitrate$ column, which is shifted forward by an offset value. Machine learning algorithms are trained to predict this $future\_bitrate$ feature, namely the future $max\_bitrate$ of the link.

The offset value along with the frequency of the data collection give the time depth of the future predictions. Different values were tested between 1 and 40. The quality of the prediction decreases significantly for an offset greater than 20. The value of 5 constitutes an optimal comprise between time and accuracy, without being too resource-intensive or time consuming. A frequency of 3 seconds with an offset of 5 means that our model will predict the $future\_bitrate$ value 15 seconds in the future. 15 seconds is enough time for a reaction mechanism to correct future congestion. Finally, the dataset is normalized between 0 and 1 using a min-max scaler. This normalization step is important for the training process, since the difference in the features scales can cause problems during the training.

%%%%%%%%%%%%%%%%%%%%%%%%%%%%%%%%%%%%%%%%%%%%%%%%%%%%%%%%%%%%%%%%%%%%%%%%%%%%%%%%%%%%%%%%%%%%%%%%%%%%%%%%%%%%%%%%%

\section{Experiments}

\subsection{Machine learning algorithms}

Three types of machine learning algorithms are tested and evaluated in this paper: the Autoregressive Integrated Moving Average (ARIMA), the Multilayer Perceptron (MLP) and the Long Short-Term Memory network (LSTM). These algorithms are trained in a supervised way and try to predict the future bandwidth utilization based on 1 feature for ARIMA ($future\_bitrate$), and 115 features for the two neural networks. The dataset was randomly divided into $k$ subsets of equal size. The first subset was used for validation and the models were trained on $k$-1 subsets ($k$-fold cross-validation). This technique gives a more accurate estimate of the models' performance by reducing bias and sample variability between training and test data \cite{Bengio:2004:NUE:1005332.1044695}. The models are trained on several links to be able to generalize their predictions to any link in the network.

\textbf{ARIMA} is a widely used approach for analyzing and forecasting time series data. If necessary, time series are made stationary by differencing or applying nonlinear transformations. A stationary time series has the property that the mean, variance, and autocorrelation are constant over time. This type of data is easier to predict since its statistical properties will be the same in the future. The ARIMA model is fitted on a single feature: the future bandwidth utilization $future\_bitrate$.

\textbf{MLPs} are a classical feedforward neural network architecture. They are comprised of one or more layers of nodes, which are fully connected. During the training phase, the output values are compared with the correct results to compute the value of the error function. This error is then fed back through the network to adjust the weights of each connection and improve its results. After a certain number of training cycles (called \textit{epochs}), the MLP converges to a solution which minimizes the value of the error function. For this work, a 116-256-128-64-1 topology is used, with ReLU as the activation function. MLPs are fast to train and suitable for regression prediction problems, particularly with tabular datasets. Hyperparameters were tuned manually, with a batch size of 128 during 10 epochs, using the Adam optimizer.

\textbf{LSTMs} are a special kind of recurrent neural networks, capable of learning long-term dependencies \cite{Hochreiter:1997:LSM:1246443.1246450}. They are designed to recognize patterns in sequence of data, with an additional temporal dimension. This feature is particularly relevant to our problem, which requires learning new network behaviors while remembering past events. Hyperparameters were tuned manually, for a batch size of 128 during 10 epochs, using the Adam optimizer. The final architecture consists of 300 LSTM units and 116 ReLu units.

These three models were chosen to assess the complexity of the problem being modelled. ARIMA is able to model periodic phenomena such as seasonal sales with good accuracy. If it proves to be effective, this will mean that the prediction of network link usage can be summarized in one feature. MLP and LSTM are two models that use all 115 features for their predictions. The difference is that LSTMs keep a memory of past events, which can be useful for predicting future bandwidth utilization. This type of neural network is typically used in time series problems. But an MLP processes data faster, which is an advantage for real-time use. This architecture will therefore be preferred to the LSTM if the results are similar.

\subsection{Model Validation and Results}

After the dataframe creation phase, one dataset is created for each of the 22 monitored interfaces. Models are validated using $k$-fold cross-validation: they are trained on $k$-1 datasets and validated on the remaining one. By shuffling the different folds, we can thus obtain all the permutations of the $k$-1 datasets and obtain $k$ evaluation scores for each model. This technique also ensures that the model is validated on an interface never seen before, which is a realistic use case.

\begin{figure}[htpb]
\centering \includegraphics[scale=0.55]{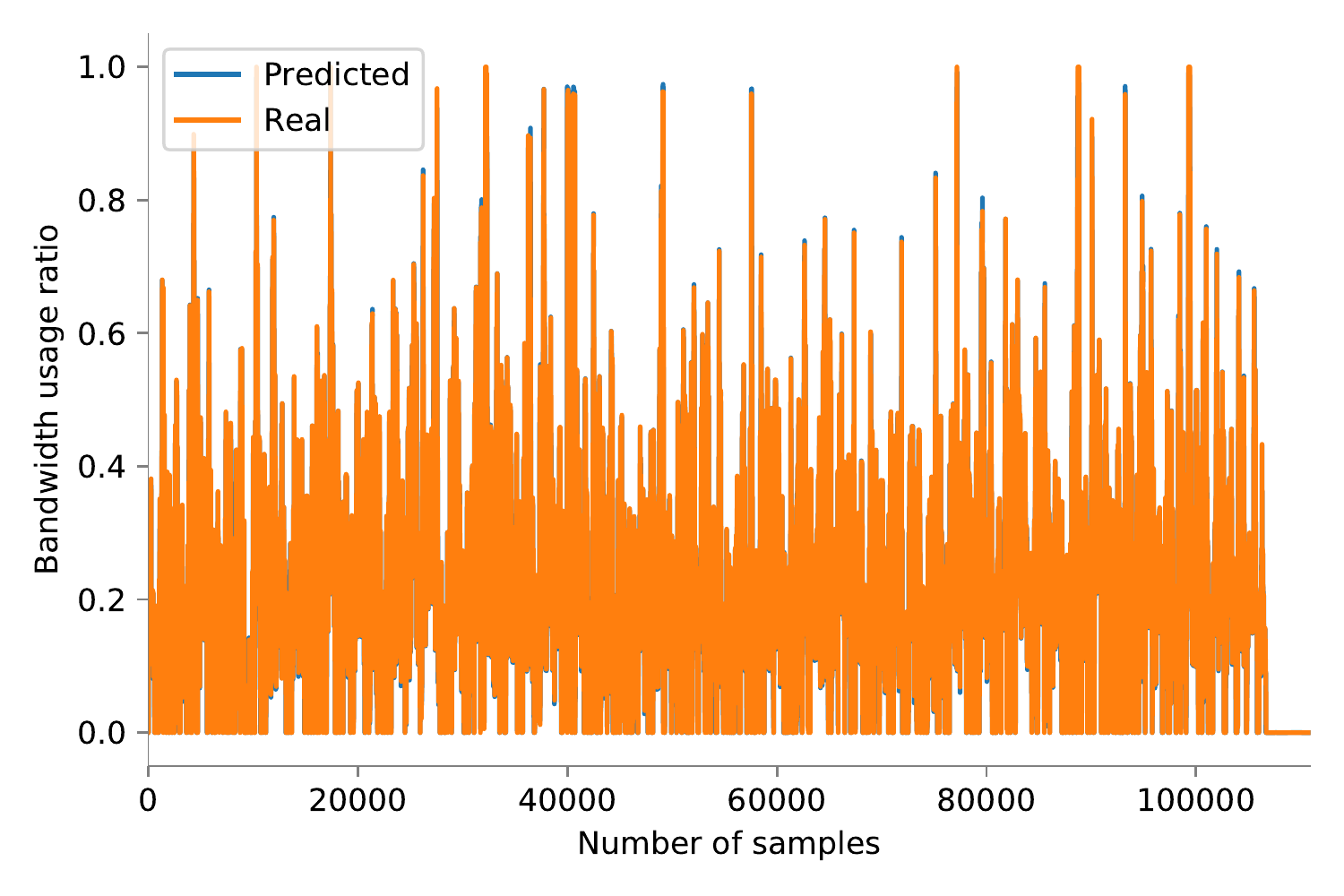}
\caption{LSTM predictions vs. actual values for one interface.}
\label{figure-lstm-predvsactual}
\end{figure}

\begin{figure}[htpb]
\centering \includegraphics[scale=0.55]{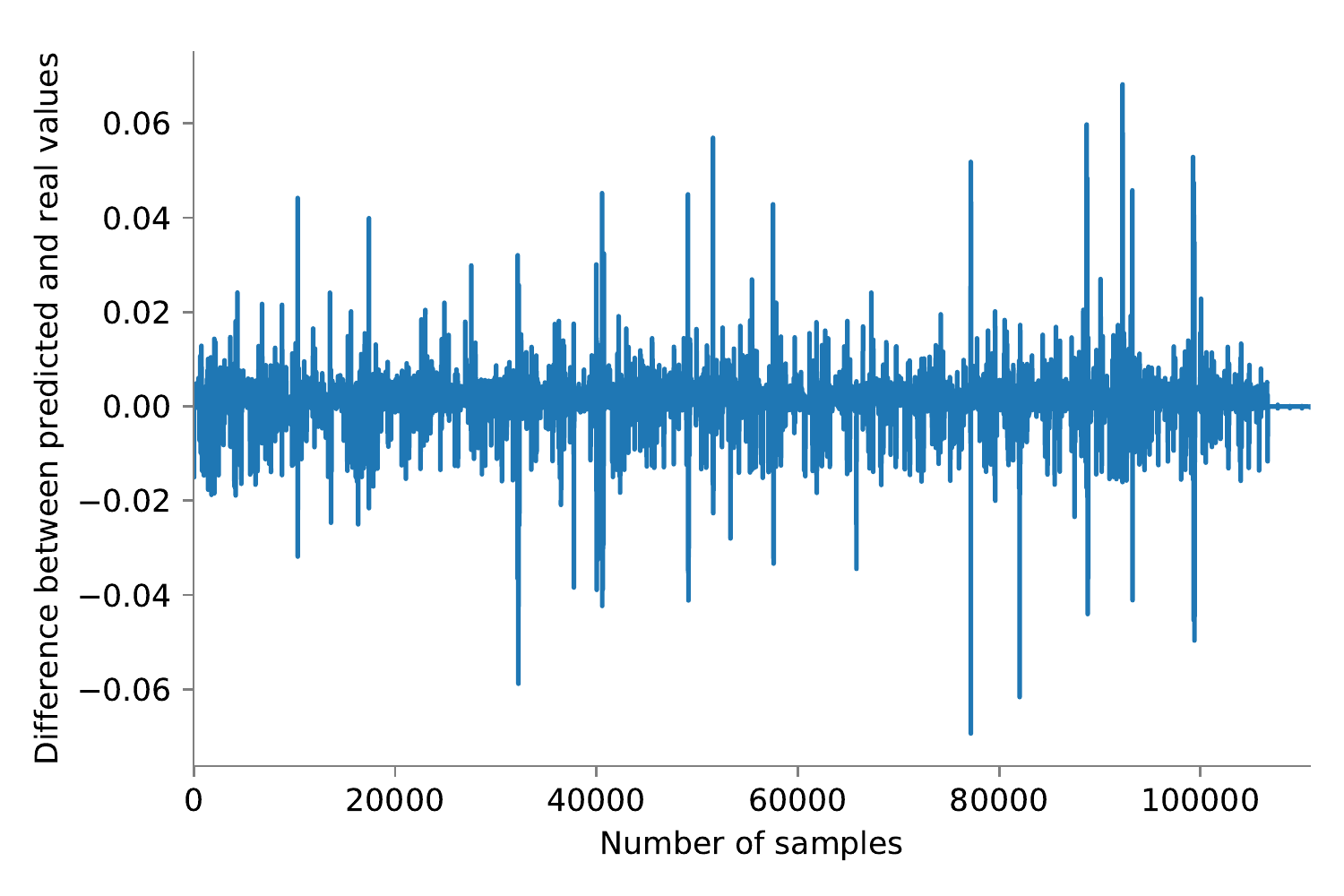}
\caption{LSTM difference between predicted and actual values for one interface.}
\label{figure-lstm-difference}
\end{figure}

\begin{figure}[htpb]
\centering \includegraphics[scale=0.55]{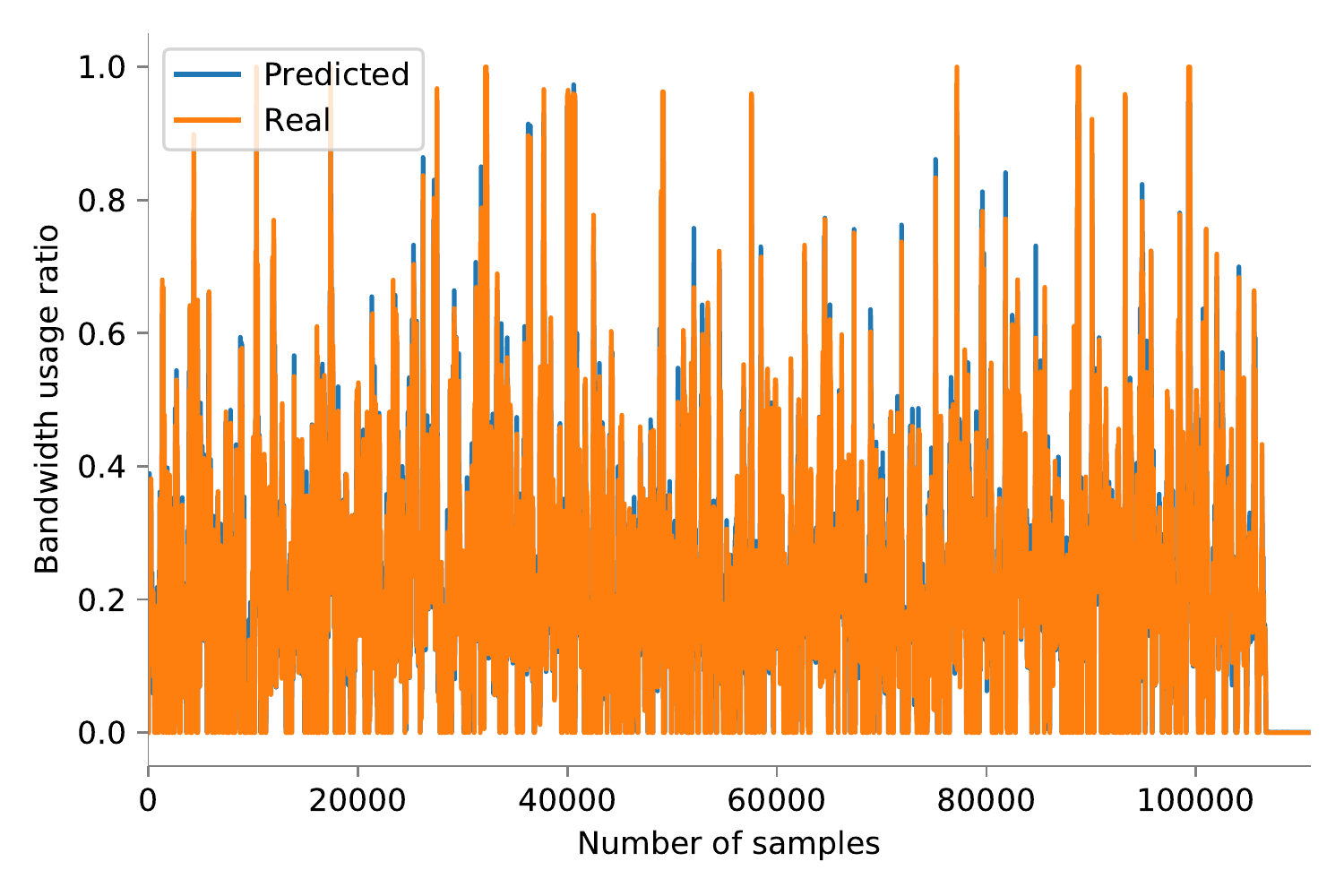}
\caption{MLP predictions vs. actual values for one interface.}
\label{figure-mlp-predvsactual}
\end{figure}

\begin{figure}[htpb]
\centering \includegraphics[scale=0.55]{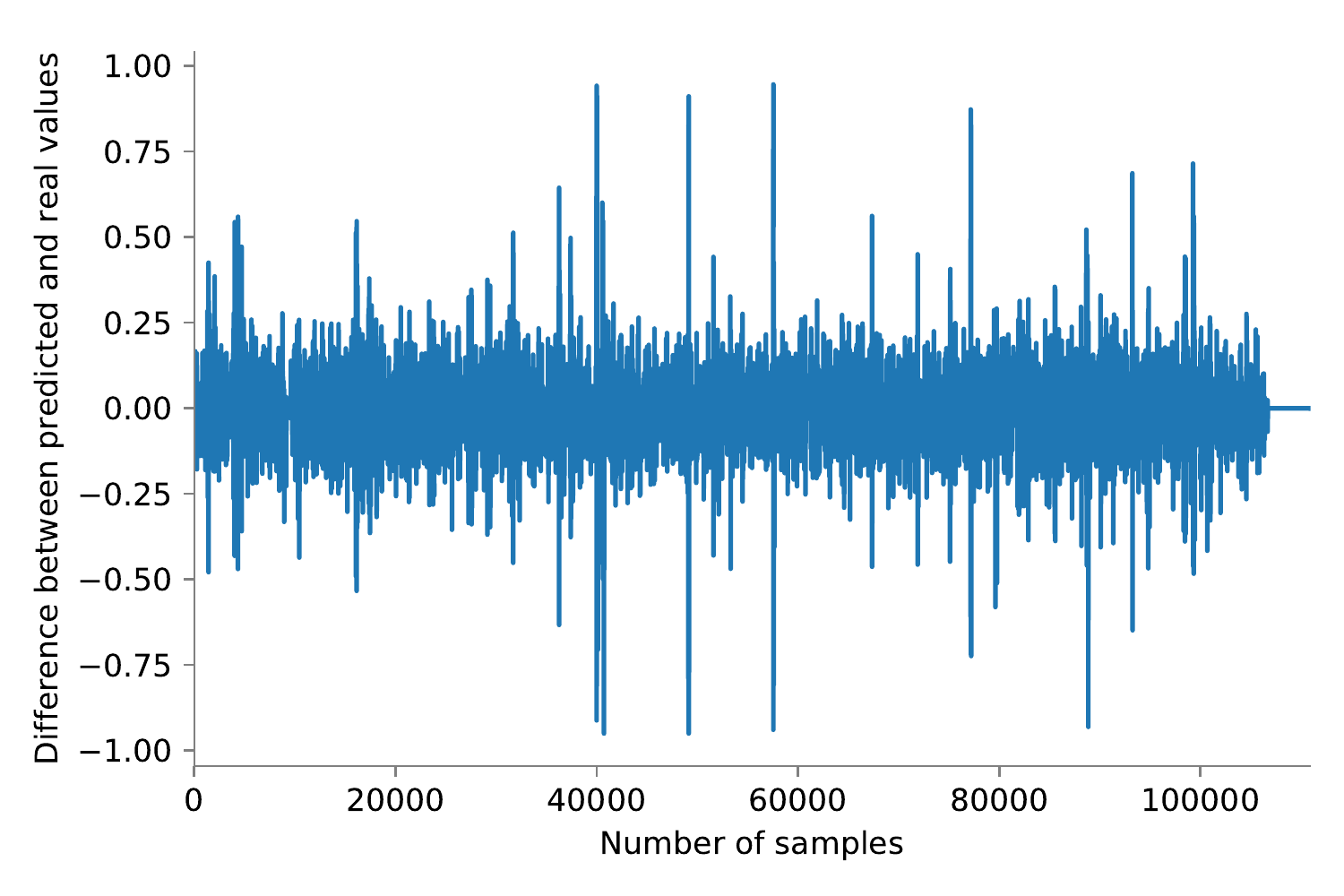}     
\caption{MLP difference between predicted and actual values for one interface.}
\label{figure-mlp-difference}
\end{figure}

Due to memory limitations and the fact that some interfaces were reciprocal (in the sense that they are directly connected to each other and the upstream traffic of one is the downstream traffic of the other), we cherry picked 8 interfaces for the evaluation. Four metrics commonly used in time series forecasting were chosen to evaluate the performance of the models: bias (systematic deviation from the actual values), Mean Absolute Error (MAE), Mean Squared Error (MSE) and Root Mean Squared Error (RMSE). Table \ref{tab:scores-table} shows the averaged values of these metrics for each model. Fig. \ref{figure-lstm-predvsactual} and \ref{figure-lstm-difference} show the predictions of the LSTM model, while Fig. \ref{figure-mlp-predvsactual} and \ref{figure-mlp-difference} show the predictions of the MLP model.

% Fig. \ref{figure-lstm-difference} and \ref{figure-mlp-difference} show, respectively, the predictions of the LSTM and the MLP models.

% Fig. \ref{figure-lstm-predvsactual} and \ref{figure-lstm-difference} show the predictions of the LSTM model, while Fig. \ref{figure-mlp-predvsactual} and \ref{figure-mlp-difference} show the predictions of the MLP model.

\begin{table}[htpb]
\centering
\caption{Averaged $k$-fold cross-validation scores}
\label{tab:scores-table}
\begin{tabular}{|c|c|c|c|c|}
\hline
\textbf{Model} & \textbf{Avg. Bias} & \textbf{Avg. MAE} & \textbf{Avg. MSE} & \textbf{Avg. RMSE}\\ \hline
ARIMA & 0.161129 & 0.162010 & 0.071085 & 0.266617\\ \hline
MLP & 0.001604 & 0.022826 & 0.002965 & 0.051939\\ \hline
LSTM & -0.002142 & 0.004272 & 0.001444 & 0.012233\\ \hline
\end{tabular}
\end{table}

It is immediately apparent from the results that ARIMA's predictions have a high bias and a poor accuracy compared to the other models. The errors produced by ARIMA are too significant for this model to be used in a real context. These poor results can be attributed to the fact that the model is fitted on a single feature. The other features could contribute greatly to the quality of the algorithm's prediction.

LSTM models perform better than MLPs. For all interfaces the difference between the actual value and the predicted one rarely exceeded 3\%, which means that the future bandwidth consumption is predicted with an error of +/- 300 Kbps on 100 Mbps links. On the other hand, the difference for the MLP sometimes exceeded 20\%, which might cause significant underestimation or overestimation of the future link capacity. The ability of LSTM models to recall past variations in bandwidth usage may explain the high quality of their predictions.

In addition, the results for some interfaces were significantly better than for others. The results were worse when the interface had higher variance of the traffic carried during the simulation phase. In other words, the predictions were much closer to the real values for interfaces that were less utilized during simulation than for interfaces that had seen the whole spectrum from 0 to 100 Mbps.

\subsection{Real-Time Prediction}

After having trained and evaluated the LSTM model, experiments with real-time bandwidth usage prediction were conducted. The objective was to integrate the real-time prediction mechanism with a Software-Defined Networking (SDN) platform in order to be able to detect future bottlenecks and react proactively to avoid them. SDN is a centrally managed network architecture, where a controller maintains a global view of the network and is able to dynamically reconfigure network devices to meet certain needs.

In this case, the SDN controller was running in the Orchestrator virtual machine and was also in charge of receiving collected data from the routers and using the trained LSTM model to predict future bandwidth consumption on their interfaces. Open vSwitch (OVS) \cite{Pfaff:2015:DIO:2789770.2789779} was installed on the router virtual machines. Open vSwitch is a virtual network switch, which uses the OpenFlow protocol to connect to the controller and receive reconfiguration commands. Rsync was used in order to provide the controller with the necessary data for the prediction. It transferred to the controller the data collected by the router every 3 seconds. The controller then feeds them to the LSTM model and obtain a future value of the bandwidth usage ratio.

In this test, we only monitored one router (namely router2) whose interfaces were bridged to the OVS switch so that the controller could have a view of the traffic passed through them. For the prediction data, as well as the OpenFlow signaling messages, we opted for an out-of-band approach. In other words, the switch-controller communication and the prediction data used a different dedicated channel and not the bridged interfaces used for application data.

\begin{figure}[htpb]
\centering \includegraphics[scale=0.50]{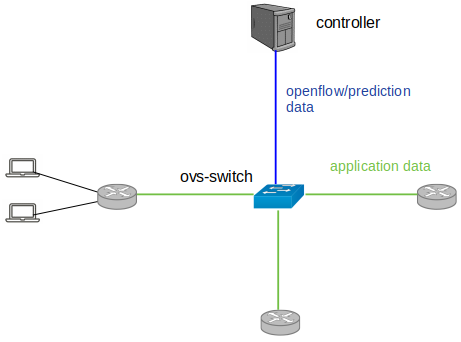}
\caption{Leveraging SDN for Proactive Management of the Network.}
\label{sdn-integration}
\end{figure}

A separate instance of the prediction mechanism was running for each one of router2's interfaces. With this setup, the controller received prediction data every 3 seconds and was able to predict each interface's bandwidth 15 seconds ahead. As a prevention measure we experimented with two solutions. A radical one was to block traffic if the bandwidth prediction exceeded a certain threshold. For example, if the prediction for an interface exceeded 0.8 (80\% of the link's capacity), the controller instructed the switch to drop new incoming packets from this interface to avoid link overload. A more sophisticated approach was load-balancing the traffic upon reaching a certain threshold. For instance, if the outgoing bandwidth reached 50\% of the capacity of a certain link, the controller installed flow rules on the switch that spread traffic across other available links.

%%%%%%%%%%%%%%%%%%%%%%%%%%%%%%%%%%%%%%%%%%%%%%%%%%%%%%%%%%%%%%%%%%%%%%%%%%%%%%%%%%%%%%%%%%%%%%%%%%%%%%%%%%%%%%%%%%%%

\section{Conclusions and Future Work}

Three machine learning algorithms (ARIMA, MLP, and LSTM) were tested to predict bandwith usage ratios 15 seconds ahead. LSTM models have shown their ability to predict network link usage with high accuracy (errors below 3\%). These models can be deployed in real time and synchronized with a SDN platform to prevent congestion before it occurs. This type of architecture allows the use of load balancing mechanisms to avoid packet loss by maximizing network throughput.

It is possible that we could get similar results (or even improve them) by removing some features, especially those collected by Dstat that do not pertain to network traffic. To that end, we intend to explore dimensionality reduction methods or simply try smaller feature sets and evaluate their performance. Less features would mean less resources required for prediction, as well as less bandwidth needed to upload the prediction data to the controller.

We also intend to investigate predictability of other parameters that might indicate congestion or other network problems. An example would be prediction of future RAM and CPU usage on the router. Resource depletion of network equipment would mean lost packets and high latency which in turn would cause congestion and degradation of the network performance in general.

Finally, we also plan to migrate to a different testbed network possibly with real hardware in order to re-evaluate the performance and make sure that our solution was not somehow topology dependent. For the same reasons, we also plan to use real applications to generate traffic like FTP, HTTP, and RTP or a real traffic generator.

%%%%%%%%%%%%%%%%%%%%%%%%%%%%%%%%%%%%%%%%%%%%%%%%%%%%%%%%%%%%%%%%%%%%%%%%%%%%%%%%%%%%%%%%%%%%%%%%%%%%%%%%%%%%%%%%%%%%

\bibliographystyle{./IEEEtran}
\bibliography{./bibfile}

% \begin{thebibliography}{00}
% \bibitem{b1} G. Eason, B. Noble, and I. N. Sneddon, ``On certain integrals of Lipschitz-Hankel type involving products of Bessel functions,'' Phil. Trans. Roy. Soc. London, vol. A247, pp. 529--551, April 1955.
% \bibitem{b2} J. Clerk Maxwell, A Treatise on Electricity and Magnetism, 3rd ed., vol. 2. Oxford: Clarendon, 1892, pp.68--73.
% \bibitem{b3} I. S. Jacobs and C. P. Bean, ``Fine particles, thin films and exchange anisotropy,'' in Magnetism, vol. III, G. T. Rado and H. Suhl, Eds. New York: Academic, 1963, pp. 271--350.
% \bibitem{b4} K. Elissa, ``Title of paper if known,'' unpublished.
% \bibitem{b5} R. Nicole, ``Title of paper with only first word capitalized,'' J. Name Stand. Abbrev., in press.
% \bibitem{b6} Y. Yorozu, M. Hirano, K. Oka, and Y. Tagawa, ``Electron spectroscopy studies on magneto-optical media and plastic substrate interface,'' IEEE Transl. J. Magn. Japan, vol. 2, pp. 740--741, August 1987 [Digests 9th Annual Conf. Magnetics Japan, p. 301, 1982].
% \bibitem{b7} M. Young, The Technical Writer's Handbook. Mill Valley, CA: University Science, 1989.
% \end{thebibliography}
% \vspace{12pt}
% \color{red}
% IEEE conference templates contain guidance text for composing and formatting conference papers. Please ensure that all template text is removed from your conference paper prior to submission to the conference. Failure to remove the template text from your paper may result in your paper not being published.

\end{document}